# Thermally-Driven Atmospheric Escape

Robert E. Johnson
   Engineering Physics, Thornton Hall B102, University of Virginia,
      Charlottesville, VA 22902
   Physics Department, NYU, NY, NY 10003
   rej@virginia.edu

**Abstract**

Accurately determining escape rates from a planet's atmosphere is critical for determining its evolution. Escape can be driven by upward thermal conduction of energy deposited well below the exobase, as well as by non-thermal processes produced by energy deposited in the exobase region. Recent applications of a model for escape driven by upward thermal conduction, called the slow hydrodynamic escape model, have resulted in surprisingly large loss rates for the thick atmosphere of Titan, Saturn's largest moon. Based on a molecular kinetic simulation of the exobase region, these rates appear to be orders of magnitude too large. Because of the large amount of Cassini data already available for Titan's upper atmosphere and the wealth of data expected within the next decade for the atmospheres of Pluto, Mars, and extrasolar planets, accurately determining present escape rates is critical for understanding their evolution. Therefore, the slow hydrodynamic model is evaluated here. It is shown that such a model cannot give a reliable description of the atmospheric temperature profile *unless* it is coupled to a molecular kinetic description of the exobase region. Therefore, the present escape rates for Titan and Pluto must be re-evaluated using atmospheric models described in this paper.

**Introduction**

There is considerable interest in descriptions of atmospheric escape due to the extensive Cassini measurements in Titan's thermosphere, the coming encounter of New Horizons with Pluto, the MAVEN mission to study the thermosphere of Mars and the evolution of extrasolar planet atmospheres. For many solar system bodies atmospheric escape is dominated by non-thermal processes induced by energy directly deposited in the exobase region, as discussed for Mars (e.g., Chaufray et al. 2007) and Titan (e.g., Johnson 2009). Escape can also occur by energy deposited well below the exobase and transported upward by thermal conduction, as described for Pluto (e.g., Hunten and Watson, 1982; McNutt 1989; Krasnopolsky 1999; Strobel 2008a). Surprisingly, analysis of the extensive Cassini-Huygens data on the density profile in Titan's upper atmosphere has lead to huge differences in the estimates of the present escape rates (e.g., Johnson et al. 2009 for a review). Prior to Cassini's tour of the Saturnian system, escape of nitrogen and methane from Titan had been assumed to occur by nonthermal processes (e.g., Shematovich et al. 2003; Michael et al. 2005). However, the density vs. altitude profiles have been recently analyzed *assuming* that escape is driven by upward thermal conduction of energy deposited well below the exobase (Strobel 2008a, 2009; Yelle et al. 2008). A model called the `slow hydro-dynamic escape' model, referred to here as the

SHE model, was applied to Titan by Strobel (2008a). This model was intended to describe cases intermediate between Jeans escape and hydrodynamic escape (e.g., Hunten 1982). It is based on solving the 1D radial fluid dynamic equations when the net upward flow velocity below the exobase is much less than the speed of sound, hence the word `slow'.

The large escape rates derived from the SHE model were suggested to be produced by distortion of the molecular velocity distribution function in the exobase region (Yelle et al. 2008; Strobel 2009). However, recent kinetic Monte Carlo simulations suggest that the nitrogen and methane escape from Titan's atmosphere, obtained using the SHE model, appear to be much too large (Tucker and Johnson 2009). Therefore, the SHE model is examined below in order to determine its general applicability and, in particular, its relevance to escape of nitrogen and methane from Titan.

**Thermal Escape**

The Jeans parameter, $\lambda$, is typically used to determine the importance of thermal models for escape. $\lambda$ is the ratio between the gravitational binding energy of a molecule in a planet's atmosphere, $\Phi_g(r)$, to its thermal energy, $kT$ : $\lambda = \Phi_g(r)/kT$. Here $k$ is the Boltzmann constant, $T$ the temperature, with $\Phi_g(r) = GMm/r$, where $G$ is the gravitational constant, $M$ the planet's mass, $m$ the molecular mass, and $r$ the distance from the center of the planet. The Jeans parameter can also be written as $\lambda = (v_{esc}/c)^2$, where $v_{esc}$ is the escape speed at $r$ and $c = (kT/m)^{1/2}$ is of the order of the speed of sound. When the net heating at depth, after accounting for radiative cooling, is high and neither downward nor horizontal heat transport can remove energy fast enough, then the temperature and pressure can increase until the gas flows outward into space removing energy. This process, referred to as hydrodynamic escape, occurs when the value of the Jeans parameter at the exobase, $\lambda_x$, becomes small (e.g., Hunten1982). It is often used to describe the evolution of young, hot atmospheres having a large escaping hydrogen component that can entrain and carry off heavier species. On the other hand when $\lambda_x$ is large, thermal escape occurs on a molecule by molecule basis, in an evaporative process called Jeans escape. The slow hydrodynamic escape model (SHE), meant to be intermediate between these, is described in a number of publications (e.g., Johnson et al. 2008 for a summary).

The mass loss rate for a single species atmosphere in the SHE model is obtained by solving 1D radial fluid dynamic equations, typically using scaled variables. These equations were initially used to describe the escape of solar wind ions assuming thermal conduction is maintained by the electrons and occurs along the radial field lines (e.g., Parker 1964). The model was subsequently applied to atmospheres containing only neutrals. It was used to estimate the outflow from Pluto driven by thermal conduction (Watson et al. 1981; McNutt 1989; Krasnopolsky, 1999; Strobel, 2008b) and was recently applied to Titan (Strobel 2008a; 2009).

From the 1D continuity equation, atmospheric escape requires a constant net flow of molecules from the lower boundary to the upper boundary: i.e., $4\pi r^2 n v = \varphi$, where $v$ is the radial flow velocity, $r$ the radial position from the center of the body, $n$ the density at $r$, and $\varphi$ the escape rate. The radial energy and momentum equations are solved in terms of an escape rate, $\varphi$. The parameter $\varphi$ is subsequently used to fit the available density data. In this model 'slow' means that $mv^2/kT$ is small below the exobase, so terms explicitly containing $v^2$ are dropped while keeping $\varphi$ fixed. Therefore, SHE is suggested to apply when $\lambda = (v_{esc}/c)^2 > \sim 10$ and in regions where $v < \sim 0.3c$ (e.g., Strobel 2008; Watson et al. 1981). The net heating rate, $Q(r)$ (heating minus radiative cooling), is used in the energy conservation equation. For a thermal conductivity $K$, heat capacity per molecule $C_p$, and gravitational energy $\Phi_g(r)$, the integrated energy equation gives the thermal flux:

$$[\phi(C_p T - \Phi_g(r)) - 4\pi r^2 (K\frac{dT}{dr})]_{r0}^{r} = \int_{r0}^{r} 4\pi r^2 Q dr \qquad (1)$$

where the $v^2$ terms are dropped.

For $n$ and T specified at a lower boundary, Eq. 1 is solved along with the momentum equation in which the terms $v^2$ are also dropped and viscosity is typically ignored. The heat flux through the lower boundary and $\varphi$ are constrained in SHE by assuming that $T \to 0$ as $r \to \infty$, and by the atmospheric density profile, as described below. Solutions for the temperature profile and escape flux have been obtained for Pluto and Titan with heat supplied by upward conduction through the lower boundary ($Q(r)=0$ in Eq. 1) or with net solar heating ($Q(r)\neq 0$) in the integrated volume.

Since $\varphi$ is constant, v increases as the density decreases, eventually approaching the speed of sound. Therefore, the energy equation is, typically, not solved to infinity (e.g., McNutt 1989). Strobel (2008a, b), for instance, solves these equations up to a region above the exobase where $v < \sim 0.3c$ and then finds solutions for which the asymptotic behavior of $n$ and T are consistent with the boundary conditions at infinity.

Although ignoring the upward flow speed below the exobase is reasonable for $\lambda > \sim 10$, continuing the 1D radial energy equation into the region above the exobase is not correct. That is, the probability of collisions decreases exponentially with radial distance rapidly becoming negligible, but the thermal conductivity used is independent of density. However, it has been argued that distortions in the tail of molecular velocity distribution near the exobase act to power the heat transport and produce escape (Yelle et al. 2008; Strobel 2009). These aspects are examined below based on kinetic theory and simulations of the Boltzmann equations, followed by a discussion of the applicability of SHE.

**Kinetic Theory**

The behavior of molecules in a planetary atmosphere can be accurately described by the Boltzmann equations (e.g., Chamberlain and Hunten 1987) in which the spatial and temporal gradients in the phase space densities are determined by the collisions between the molecules and by the forces acting on the molecules. Taking the five principal moments of these equations for each species (e.g., Chapman and Cowling 1970) results in the fluid dynamic equations: the continuity, energy, and momentum equations for the molecular density, energy density, and mean flow velocity of the molecules. It is the 1D radial versions of these equations that are used in SHE. These apply in regions of the atmosphere in which the mean free path for collisions, $\ell_c$, is much less than the scale for significant changes in atmospheric properties, typically described by the local scale height of atmosphere, $H$. The ratio, $\ell_c/H=Kn$, is called the Knudsen number. Therefore, for $Kn \ll 1$, the velocity distribution function (VDF) for the molecules is well described by a Maxwell-Boltzmann (MB) distribution and the fluid dynamic equations can accurately describe the behavior of the gas.

Gradients in these moments, which are locally averaged quantities, result in the thermal conduction, viscous and diffusion terms. These account for the flow of molecules between two adjacent volumes of gas, both assumed to have MB VDF, but with slightly different temperatures, flow speeds or compositions. The transport of energy and momentum between neighboring volumes depends on the molecular speeds. Prior to making a collision, more molecules will flow from the hotter region than from the colder region, and those in the tail of the hotter region will be lost to the neighboring volume faster. Therefore, in addition to the transfer of heat and momentum, the MB VDF is *distorted* by an amount that depends on the size of the gradients and on the mean free path, and this distortion is largest in the tail of the distribution where the velocities are highest. When $Kn \ll 1$, the distortion of the VDF is typically ignored and considered a second order effect, and the transport of energy, momentum and mass is well described by the thermal conduction, viscous and diffusion terms. But the presence of such gradients implies the VDF have a slightly non-Maxwellian character even when $Kn$ is small.

The distortion of the VDF can become significant near the exobase affecting the transport properties and estimates of the escape flux. The effect of the distortion can be calculated using higher order moments of the Boltzmann equations, such as the 13 moment equations (e.g. Chapman and Cowling 1970, Hirschfelder et al. 1964; Schunk and Nagy 2000). These are used in Cui et al. (2008) and Yelle et al. (2008) for $H_2$ escape from Titan's upper atmosphere. Since the first order perturbations to the VDF come from gradients in temperature and flow speed, the thermal conductivity and viscosity determine the size of the distortion. A simple estimate of the change in the VDF, $f(w)$, in the $z$ direction, with $w$ the $z$ (upward) component of the molecular velocity, due to a small vertical temperature gradient is $f(w) \sim f^{(0)}(w) - \ell_c (dT/dz) df^{(0)}/dT$. Here $f^{(0)}$ is the MD VDF for a temperature $T$ (e.g., Chapman and Cowling 1970). Since the viscous effects are typically assumed to be small in SHE, only the temperature gradient is considered here. Writing $df^{(0)}/dT = (f^{(0)}/2T)[mw^2/kT - 1]$, it is seen that for large $w$, $f$ can be significantly affected, as discussed above.

If there is no external heat source for the exobase region, but escape is occurring, then the heat removed by the exiting molecules leads to a negative temperature gradient in the exobase region, a process referred to as adiabatic cooling. In this case the exobase temperature, $T_x$, is lower than $T$ below the exobase so that, in expression for $f(w)$ above, molecules with escape energies are replenished from below refilling and adding to the tail of the VDF. Based on this, it is argued that the presence of the temperature gradient 'enhances' Jeans escape for $CH_4$ and $N_2$ (Yelle et al. 2008; Strobel 2009). What is meant is that the escape flux is enhanced over the Jeans rate calculated using the exobase temperature, $T_x$. Since $T_x$ is lower than $T$ below the exobase due to escape, this is somewhat circular. But a rough estimate of the enhancement is readily obtained. Since the molecules which populate the hot tail in the exobase region come from below, their contribution to escape can be estimated using the VDF below that altitude from which hot molecules can directly escape to space, a few mean free paths below the exobase (see e.g., Johnson et al. 2008). For $\lambda_x > \sim 10$ the tail of the distribution dominates escape, so that the 'enhanced' escape rate is, roughly, the Jeans rate calculated using the $T$ a few scale heights below the exobase. The Jeans expression for the net escape in a 1D atmosphere is: $\varphi_J = 4\pi r_x^2 (n_x \overline{v_x}/4)(\lambda_x + 1)\exp(-\lambda_x)$ ; with $\overline{v_x} = (8kT_x/m\pi)^{1/2}$ and the subscript $x$ implies values at the nominal exobase. Replacing their exobase temperatures, $T_x$, from Yelle et al. (2008) and Strobel et al. (2008) by their values of $T$ at a few scale heights below the exobase, the `enhancement' in Jeans rate for escape of $N_2$ ($\lambda_x \sim 40$) or for $CH_4$ ($\lambda_x \sim 20$) from Titan is small and is orders of magnitude below the rates calculated by these authors. This result is confirmed by direct simulations of the kinetics in the exobase region as discussed below.

**Monte Carlo Simulations**

Monte Carlo simulations can b used to describe the kinetic behavior of a gas. Since such simulations are equivalent to solving the Boltzmann equations (e.g., Bird 1994), they reproduce the results of the fluid dynamic equations well below the exobase where $Kn < \sim 0.1$. In addition, such simulations are often more readily implemented than the Boltzmann equations. In Monte Carlo simulations, the atmospheric density is described by a large number of representative molecules. Collisions occurring between these representative molecules are described by the cross sections for real molecules and, hence, can be calculated as accurately as necessary. Between collisions, the representative molecules move subject to the gravitational force of the planet/satellite and any neighboring body. If ion motion is considered then the electromagnetic forces can be included (Tseng et al. 2009). Since a finite volume is simulated, appropriate boundary conditions are applied: escape across some upper boundary, absorption or supply of molecules at a lower boundary, etc.

Starting with some initial distribution of molecular positions and velocities, and prescribed boundary conditions, the evolution of an atmosphere is simulated. For fixed external conditions, the simulations are run until the distributions of position and velocities of the representative molecules yield steady-state densities and temperatures. The the number of molecules entering and leaving the simulated volume is also recorded. From the position and velocities of the representative molecules, the steady state spatial distributions and VDF are calculated. Because the cross sections, forces, and boundary conditions can, in principal, be described as realistically as one likes, the accuracy is usually determined by the number of representative particles and the method for deciding when a collision occurs. One such model is the so-called Direct Simulation Monte Carlo (DSMC) model (Bird 1994) which has been applied extensively.

Unlike the fluid equations, such simulations can, in principal, describe the behavior and escape of molecules in the exobase region from $Kn \ll 1$ to $Kn \gg 1$. In practice, consideration is given to those aspects that must be well represented. For escape from Titan, the tail of the VDF above the exobase must be accurately simulated requiring a large number of representative molecules. Accuracy can also be increased by the use of adaptable grids and weights. Since the molecules well above the exobase move in large ballistic trajectories, an upper boundary is typically chosen for which the collision probability drops below some prescribed level. At this boundary the representative molecules are tested to see if they would escape or return to the atmosphere. Therefore, the grid size, weights, boundaries, and numbers of representative molecules are optimized. These procedures are well understood, and have been applied, for example, to cometary coma (e.g., Tenishev et al. 2008) and to escape from Titan induced by both the incident plasma (e.g., Shematovich et al. 2003; Michael et al. 2005; Michael and Johnson 2005) and thermal conduction (Tucker and Johnson 2009).

Simulations for a pure $N_2$ atmosphere and an $N_2$ plus $CH_4$ atmosphere were carried out to test the continuum models for escape from Titan. Values of temperature, density and heat flux from Strobel (2008a) and Yelle et al. (2008) were used at a lower boundary a few scale heights below the exobase (Tucker and Johnson 2009). These simulations were carried out in both 1D rectangular and 3D spherical coordinates to test escape driven by heat transported from below, ignoring non-thermal processes. For comparison with the 1D radial SHE model, the 3D spherical simulations were performed for an isotropic atmosphere in which the VDF, density and temperature exhibit only radial variations. Sensitivity tests were made on the cross section models, the upper and lower boundaries positions, and the number of representative molecules. The simulations for the $N_2$, as in Strobel (2008b, 2009), and for the $N_2 + CH_4$, as in Yelle et al. (2008), were reported. It was found that, due to the large value of $\lambda_x$, the radial and rectangular simulations results were not very different until well above Titan's exobase. More importantly, it was shown that the 1D continuum models, when extended above the exobase, overestimated the escape rates by orders of magnitude, consistent with the discussion above. Therefore, the continuum estimates of escape at Titan appear to be incorrect, so that care should be taken in applying SHE to other bodies (e.g., Tucker et al. 2010).

**Applicability of SHE**

The SHE model can be understood using an analytic solution to Eq. 1. For an assumed constant $K$ with the heat deposited below the lower boundary, $r_0$, so that $Q(r > r_0) = 0$, and applying the boundary condition ($T \rightarrow 0, r \rightarrow \infty$), one obtains:

$$C_pT(r) = [1-exp(-r_\varphi/r)][-\Phi_g(r_\varphi) + \overline{E\varphi}/\varphi] + \Phi_g(r) \qquad (2a)$$

Here $r_\varphi = [\varphi C_p/4\pi K]$ is a length scale indicating when energy transport changes from pure thermal conduction, $\varphi=0$, to heat transport and escape by molecular flow. The quantity $\overline{E\varphi}$ is the energy per unit time flowing through the system from the lower boundary, so that the energy per escaping molecule transported upward from below is $\overline{E\varphi}/\varphi$. For transparency, the scaling of the variables typically carried out in SHE is not used (appendix). Setting $T = T_0$ at the lower boundary $r_0$, $T(r)$ is obtained from Eq. 2a:

$$[\Phi_g(r) - C_pT(r)] / [\Phi_g(r_0) - C_pT_0] = [1 - exp(-r_\varphi/r)] / [1 - exp(-r_\varphi/r_0)] \qquad (2b)$$

In published applications of SHE, $K$ is usually temperature dependent, but the change in $K$ is relatively small over the exobase region. In addition, the solution is typically terminated at some point above the exobase where v/c is still small, as discussed earlier, *but* with the requirement that $T(r)$ exhibits the asymptotic behavior ($T \rightarrow 0, r \rightarrow \infty$). However, the analytic expression in Eq. 2b is representative of the applications of SHE to planetary escape.

It is seen in Eq. 2b that the escape rate, $\varphi$, is a free parameter. It is not directly determined by the energy transport unless there is direct measure of the temperature profile. As the viscosity is typically estimated to be small, the hydrostatic approximation is used in SHE to determine the density, $n$: $d(nkT)/dr = n \, d(\Phi_g(r))/dr$. Inserting the $T(r)$ in Eq. 2b, and writing $x=r_\varphi/r$ and $x_0=r_\varphi/r_0$, one finds:

$$n(r) = n_0[T_0/T]\exp\{-(C_p/k)\int_x^{x_0} dx'/[x'-x_0(1-\exp(-x'))\beta_0]\} \qquad (3)$$

where $\beta_0 = [1 - (C_p/k)/\lambda_0] / [1-\exp(-x_0)]$, with $\lambda_0$ the value of the Jeans parameter at $r_0$. Since the atmospheric temperature is generally not directly measured, the hydrostatic approximation is also used to extract the atmospheric temperatures. Therefore, the value of $\varphi$ can be determined by fitting $n(r)$ to available density data below the exobase or to the temperature profile extracted from that density data.

Parameters at Titan's exobase vary with solar conditions in Strobel (2009). Representative exobase values are: $r_x \sim 4000 km$; $K_x \sim 1.4 \times 10^3\ erg cm^{-1} s^{-1} K^{-1}$; $\lambda_x \sim 40$ for $N_2$ or $\sim 20$ for $CH_4$; $T_x \sim 150 K$; and $n_x \sim 3 \times 10^7 N_2/cm^3$ with $CH_4$ less than 10% of the total in the exobase region. Using the above $K$, the length scale in Eq. 2b is $r_\varphi \sim \varphi(3 \times 10^{-24}\ km\ s)$. Substituting the Jeans value for $\varphi$ based on these parameters results in a negligible length scale: $r_\varphi \sim (3 \times 10^{-11} - 10^{-3})\ km$, depending on whether one assumes nitrogen or the methane is the predominant escaping molecule. For these very small values of $r_\varphi$, the temperature and density profiles are determined only by the thermal conductivity and the upper boundary condition: $T(r) \sim T_0 (r_0/r)$ and $n(r) \sim n_0[r_0/r]^{\lambda_0 -1}$. Such a profile is not representative of Titan's atmosphere. Using instead $K \propto T^s$, $T(r)$ would decay somewhat more slowly, $T(r) \sim T_0(r_0/r)^{-1/(1+s)}$, but the resulting radial profiles are still not characteristic.

Therefore, unless one assumes that $\varphi$ is many orders of magnitude larger than the Jeans rate for these $\lambda_x$, the SHE model *cannot* give realistic density profiles. For example, Strobel (2008) finds a mass loss rate that is relatively large even for the case $Q = 0$ above $r_0 = 3450 km$: i.e., $m\varphi \sim 2 \times 10^{28} amu/s$. The recommended value in Strobel (2009) is similar, $m\varphi \sim 3 \times 10^{28} amu/s$, so that $r_\varphi \sim 200 km$. These large rate are required to give a realistic radial dependence for the atmosphere. The corresponding large upward flux of heat is seen to cause $T(r)$ to decay *more slowly* over the exobase region than for the small $\varphi$ result above. This is the opposite to what is suggested in discussions of the model. That large $\varphi$ are favored in the SHE model is also consistent with the scaling procedure typically used, in which the energy equation (Eq.1) is divided by $\varphi$ before solving (appendix).

As stated earlier, for $\lambda_x >\sim 10$, and/ or in regions where $v <\sim 0.3c$, Eq. 1 can describe the temperature dependence of the atmosphere a few scale heights below the exobase. Therefore, rather than solving up to some region above the exobase, and forcing the temperature to go to zero at very large $r$, the region above $Kn \sim 0.1$ should be described by a molecular kinetic model. For this reason we developed a hybrid model in which the SHE equations are solved, using an estimate of the escape rate up to a radial distance $r_u$ at which $Kn \sim 0.1$. These results are then used as a lower boundary condition for a DSMC simulation describing the region from $r_u$ up to an altitude well above the exobase where $Kn >>1$. The improved estimate of the escape rate so obtained is then used to obtain a new solution of the radial 1D SHE equations. This iterative procedure was applied to the atmosphere of Pluto, a body much smaller than Titan, for which the thermal escape flux is expected to be significant. To avoid questions of the accuracy of the description of $Q(r)$, the model was applied to the case in which all the solar energy is deposited below the lower boundary: i.e., $Q = 0$ in Eq. 1. The $T(r)$ profile was found to be very different from that in Strobel (2008a) and the atmospheric loss rate was orders of magnitude smaller (Tucker et al. 2010). Of course, for very large heating rates, large escape rates driven by upward thermal conduction are possible, as will be described in subsequent work.

One reason the SHE model disagrees with the DSMC simulations is that the form for the thermal conductivity, $K$, is kept constant above the exobase. That is, in calculating $K$, the product $n\ell_c$ is typically independent of density in the fluid regime. The conductivity term in the Eq. 1 requires that collisional energy transfer occurs over distances very small compared to the scale of the density gradients: i.e., $Kn \ll 1$. This could be roughly circumvented in an ad hoc manner by replacing $n\ell_c$ by $\sim n\ell_c/(1+Kn)$ in the transition region. Such a substitution would allow the effective K to decrease with decreasing density when the mean free path length for neutral-neutral collisions is longer than the scale height.

The SHE model can also be applicable if thermal conduction occurs by means other than by neutral-neutral collisions. If, for example, ions are dragged out of an atmosphere along magnetic field lines, which is in fact the case at Titan, then collisional heat transport can effectively persist well above the exobase defined by the neutral-neutral collision cross section. This is a process we have described as 'back sputtering' (see e.g., Johnson et al. 2008; Johnson 2009).

In SHE, the requirement that T goes to zero at infinity is also problematic as seen in the analytic model above. This requires that the thermal energy of the escaping molecules is fully converted into flow energy. This assumption is critical, since it is the flow energy in SHE, and not the tail of the random thermal motion, that carries the molecules over the gravitational barrier. However, when significant escape occurs, DSMC simulations show that the radial and horizontal temperatures can diverge significantly with increasing altitude above the exobase (Tucker et al. 2010).

If the rate of heat deposition in an atmosphere is very high, and downward thermal conduction, radiation to space, and horizontal transport are inefficient, then the atmospheric T will increase until the upward flow is such that molecular escape can carry off the excess energy (e.g., Hunten 1982). But this does not imply that the SHE model is required. In the absence of direct energy deposition in the exobase region, DSMC simulations show that even for an artificially high $T_x$ at Titan ($\lambda_x \sim 10$; Tucker and Johnson 2009), the relatively large escape rate does not differ significantly from the Jeans rate. These results are consistent with the criterion suggested much earlier (e.g., Hunten 1982): if $\lambda$ approaches $\sim 2$ above the exobase, then the escape rate is not too different from the Jeans rate, but if $\lambda$ approaches $\sim 2$ near or below the exobase, then direct outflow can occur and the atmosphere is highly extended approaching a cometary description: i.e., $\ell_c \sim r_x$. The loss from Titan of the trace species, $H_2$, with $\lambda_x \sim 3$ at the exobase, is borderline (e.g., Cui et al. 2008). However, the resulting $H_2$ flux is not sufficient to drive off the heavy species, but acts to extract heat from the background molecules. For the heavy species, $N_2$ and $CH_4$, with larger $\lambda_x$ solutions to the radial 1D equations below the exobase and should be attached to a DSMC model of escape, as discussed above, or to a value of the escape rate close to the Jeans rate.

**Summary**

Detailed in situ data on the structure of atmospheres in the exobase region from spacecraft are now available for some solar system bodies, will soon be for a number of others, and are of considerable interest to modeling the evolution of exoplanet atmospheres. In this region, also called the transition region, it is critical to use models that can accurately describe the molecular kinetics. Although escape driven by thermal conduction can compete with non-thermal escape processes (e.g., Johnson et al. 2008; 2009), it is shown that the 1D, slow hydrodynamic escape model, as presently applied, can not be relied on to give accurate molecular escape rates. Although the escape rates at Titan *could* be significantly different from the Jeans rate due to nonthermal processes (e.g., DeLaHaye et al. 2007; Johnson et al. 2009), extending continuum models, such as the SHE model (Strobel 2008a, 2009) or a diffusive separation model (Yelle et al. 2009), above the exobase can give escape rates for heavy species that are orders of magnitude too large. That such escape rates for Titan are incorrect is argued here based on kinetic theory and the nature of the solutions to the SHE mode, and it has also been suggested based on simulations of the gas kinetics (Tucker and Johnson 2009). It is shown here that the 1D SHE model, as applied, appears to favor large escape rates, and, therefore, it cannot be used to readily explore the parameter space if the loss rates are unknown. This model can, in principal, be used in an iterative procedure in which the radial fluid equations are solved below the exobase and are used with an estimate of the escape rate based on Jeans escape or are attached to a molecular kinetic model, such as the DSMC model (Tucker et al. 2010). In addition, if there is a significant upward flux locally, a 3D model allowing horizontal transport is likely required. However, whether 1D or 3D continuum models are used up to the exobase, obtaining the thermal and nonthermal escape rates requires the use of a kinetic description of the gas in the transition region.

**Acknowledgement**

This work is supported by a grant from NASA's Planetary Atmosphere's Program and by a NASA Cassini Data Analysis Grant.

**Appendix**

For clarity, a standard notation was used above. The notation typically used in the SHE model is: $F=\varphi/4\pi$ is used instead of the net escape rate $\varphi$ and the variables in Eq. 1 are replaced by scaled variables: $\lambda = \Phi_g/kT = GMm/rkT$, $\zeta = Fk/K_0 r_0 \lambda_0$, $\psi = mv^2/kT_0$, $\tau = T/T_0$, and $c = (kT_0/m)^{1/2}$, where the subscript o implies the value at the lower boundary. The thermal conductivity $K$, often written as $\kappa$, can have a simple temperature dependence, $K = K_0(T/T_0)^s$, and the symbol $U = (2kT/m)^{1/2}$ is often used as the most probable speed in a Maxwellian distribution. Scaling is then carried out by essentially dividing Eq. 1 by $\varphi$, replacing r by $\lambda$ and using the variables above in

$$[(C_p T - U(r)) - 4\pi r^2 (K \frac{dT}{dr})/\phi]_{r_0}^{r} = \phi^{-1} \int_{r_o}^{r} 4\pi r^2 Q dr.$$

This results in the thermal conduction term being very large if $\varphi$ approaches to the Jeans rate and $\lambda_o \sim 20 - 40$, as it is at Titan for $CH_4$ and $N_2$.